# Pixel identification in an image using Grover Search Algorithm

**Mohd Hussain Mir[1],** Harkirat Singh[1]

[1]Deptt of Physics National Institute Of Technology Srinagar, Hazratbal Srinagar, Jammu and Kashmir, India, 190006.

**Email:** hussain_19msc19@nitsri.net

## Abstract

Quantum Computing offers an entirely new way of doing computation governed by the rules of quantum mechanics like Superposition and Entanglement. These rules allow us to do computation over all the possible states simultaneously. Hence, offering exponentially higher computation power than the present classical computers. Quantum computing algorithms are entirely different from classical algorithms due to quantum parallel computing derived from quantum state superposition and entanglement, has natural advantages over classical image processing. Grover algorithm is a quantum based search algorithm used to find the correct answer from an unsorted database by computing on all the inputs simultaneously. Thus, giving us a quadratic speed-up of order $O(n)^{1/2}$ in comparison to the classical algorithm which offers speedup with order $O(n)$.

We used Grover algorithm for identifying the black pixel in a (2x2) classical image by first converting it into a quantum state and then running the Grover algorithm for identifying the pixel with 0 value maximum gray-scale intensity. This technique has applications in areas like steganography offering data encryption between users, image segmentation etc.

# 1 Introduction

The computational advantage offered by quantum computers over classical computers[1], using the laws of quantum mechanics like superposition and entanglement has gained researcher's interest in recent years. Image processing plays a vital role in all branches of the military and commercial sectors. There are many image processing areas like surveillance, target tracking navigation, and medical sciences where image processing plays a vital role. However, due to the limited computational power of classical computers, some processing tasks with a large set of possible models can't be achieved by classical computers as they require very high computational power that is only possible with quantum computers[2]. Steganography[3] is the process in which some highly confidential data is hidden in some form in an image which could be any ordinary image like a cat or a tree. The image used in this process is called stego-image. This process doesn't change any visible feature in an image. Thus, any intruder who wants to get to the data doesn't even know that data is hidden in it[4]. The data can only be accessed by a person having a specific key while it doesn't even exist for any other person. Data security is very important when it comes to information sharing between users. Our motive is to hide the data into an image and retrieve it using a search algorithm. In quantum image processing[5], quantum image representation plays a key role, which determines the kinds of processing tasks and how well they can be performed[6]. Various methods of image representation are Qubit Lattice[7], Entangled image[8], Real Ket[9], Flexible Representation of Quantum images (FRQI)[10], Novel enhanced quantum image representation (NEQR)[11]. FRQI uses a normalized superposition to store all the pixels in an image, the same operations can be performed simultaneously on all pixels, and therefore FRQI can facilitate the computation problem of image-processing. However, the limitation of FRQI is that it uses only a single qubit to store the gray-scale information for each pixel in an image, some digital image-processing operations, for example, certain complex color operations, cannot be done based using FRQI.

The NEQR model uses the linear independent basis state of a qubit sequence to store the gray-scale value of every pixel. Therefore, to store the digital image using quantum mechanics, two entangled qubit sequences are used in NEQR, which represents the gray-scale and position information of all pixels in the image. In FRQI representation gray-scale information about the image is encoded using a single qubit while as in NEQR the grayscale information is encoded in the basis qubit states, as each basis qubit state is linearly independent the image processing task becomes much easier than FRQI. Also, the time complexity for preparing the NEQR quantum image exhibits an approximately quadratic decrease in comparison to FRQI. Grover algorithm[12] is a quantum-based search algorithm used to find the solution to a problem from an unstructured database. Classically, searching for a particular entry that is required from a database containing N unstructured entries takes on the order of O (N) checks. With Grover's algorithm, the search is performed over an index to the database elements rather than the elements themselves. The oracle in Grover's algorithm is typically thought of as a black box with the capability of recognizing the solutions satisfying user-defined constraints. The black box cannot calculate solutions, but it can verify that whether a solution is contained in the input list or not. For any given index, the oracle, f (x) returns 1 if the index is a solution to the search i.e it satisfies the user-defined constraints otherwise, it returns a 0 signifying that the index does not point to a solution[13].

In this research article two methods for image processing are used section 2 deals with NEQR processing. The NEQR method which is an enhanced version of FRQI is being used to convert the classical (2x2) gray-scale classical image into a quantum image and then the darkest pixel i.e.

minimum intensity pixel is located by using the Grover search algorithm[11]. In Section 2, image processing is done using Qiskit Jupyter environment. In Section 3, image processing is done using the Google Cirq environment. The minimum intensity pixel in both the protocols is located using the Grover search algorithm. The advantage of encoding the data into a quantum image is that the final state is the superposition of all the possible states and the computation is done over all the possible states simultaneously.

## 2(a). NEQR using Qiskit

Qiskit[14] is an open-source python framework provided by IBM used for manipulating, writing Quantum programs and then implementing them on a real Quantum computer or simulators provided at IBM quantum experience site[15].

## Computational details

The color scheme in images comprises of three intensities values known as RGB value of an image, the intensity of each color can vary from 0 where 0 means back and 255 means white. To encode each intensity ($2^q=255$) where q is the number of qubits needed to encode various intensities of a particular color and for encoding the position we need another set of qubits. Since we will be representing a two-dimensional (2x2) pixel image, we will define the position of the image by its row and column, Y, X, respectively and color by [16]

$$f(Y,X)=C^0_{yx},C^1_{yx},C^2_{yx}\ldots\ldots C^{q-2}_{yx},C^{q-1}_{yx} \in [0,1], f(Y,X) \in [0,2^{q-1}]$$

where $C^0_{yx},C^1_{yx},C^2_{yx},\ldots\ldots C^{q-2}_{yx},C^{q-1}_{yx}$ describes the color setting of |YX⟩ pixel. For example for the minimum intensity pixel located at |00⟩

$$C^0_{00\ s}= C^1_{00} = C^2_{00} = C^{q-2}_{00} = C^{q-1}_{00}= 0$$

So two additional qubits are required to encode the position information. Thus (q+2) qubits are required in total. In the table the first column represents the pixel position of the 2×2 image. In column II the binary string used to encode the color of the corresponding pixel is shown. In column III the intensity values are shown:

| Pixel position | Binary string | Grayscale intensity |
|---|---|---|
| \|00⟩ | \|00000000⟩ | 0−Black |
| \|01⟩ | \|01100100⟩ | 100−Darkshade |
| \|10⟩ | \|11001000⟩ | 200−Lightshade |
| \|11⟩ | \|11111111⟩ | 255−White |

Table representing the intensity values and the corresponding binary string needed to encode it

The steps involved in image processing using NEQR are as follows:

**Step 1:** Generate a classical (2x2) grayscale image using python.

**Step 2:** The quantum circuit of the image is formed based on the color setting of the image using NEQR code[17]. This step involves the storage of classical data as a quantum state as shown in fig.1(a).

**Step3:** Grover search algorithm[18] is then used to search for darkest pixel with intensity 0 among all the four pixels which in our case is the first pixel located at position $|00\rangle$

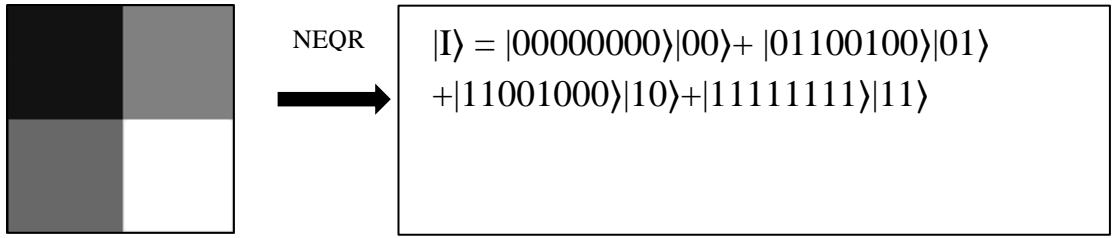

**Fig.1** A (2x2) classical image and its representation NEQR.

## 2(b). Matrix method using Google Cirq

Cirq is a python based environment provided by Google, used for writing, optimizing quantum circuits, and running them against Quantum Computers and simulators[19].

### Computational details

The image processing in Google Cirq framework is done as follows:

**Step1:** A random 2x2 pixel gray-scale image is generated in python.
**Step2:** Classical image is then converted into a 2x2 matrix in which the elements represent the grayscale intensities of the pixels at particular locations.
**Step3:** 2 qubit Grover algorithm[20] is then implemented on the matrix elements with winning state determined as the pixel with the minimum intensity such that the Grover algorithm output's the indices of the required pixel.

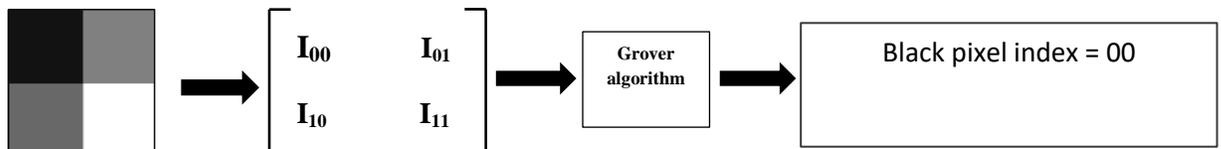

**Fig.2** Generating matrix array from image and then applying Grover algorithm to it.

## 3. Results and Discussion

### (a) Results based on Qiskit

The code is run in Python's Jupyter notebook and then on Quantum simulators provide by IBM. A 2x2pixel classical image as shown in fig.1(a) with a gray scale value ranging from 0-255 is encoded into a Quantum image circuit using NEQR code. The circuit that converts the image data into Quantum data is shown in fig.1(b). The value setting of all four pixels are separated by using barriers (shown by dotted vertical lines)

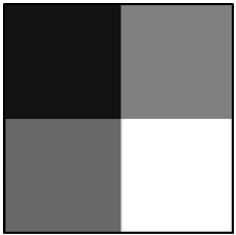

$$|I\rangle = |00000000\rangle|00\rangle + |01100100\rangle|01\rangle + |11001000\rangle|10\rangle + |11111111\rangle|11\rangle$$

**Fig.1(a)** A (2x2) classical image with minimum intensity pixel at position $|00\rangle$

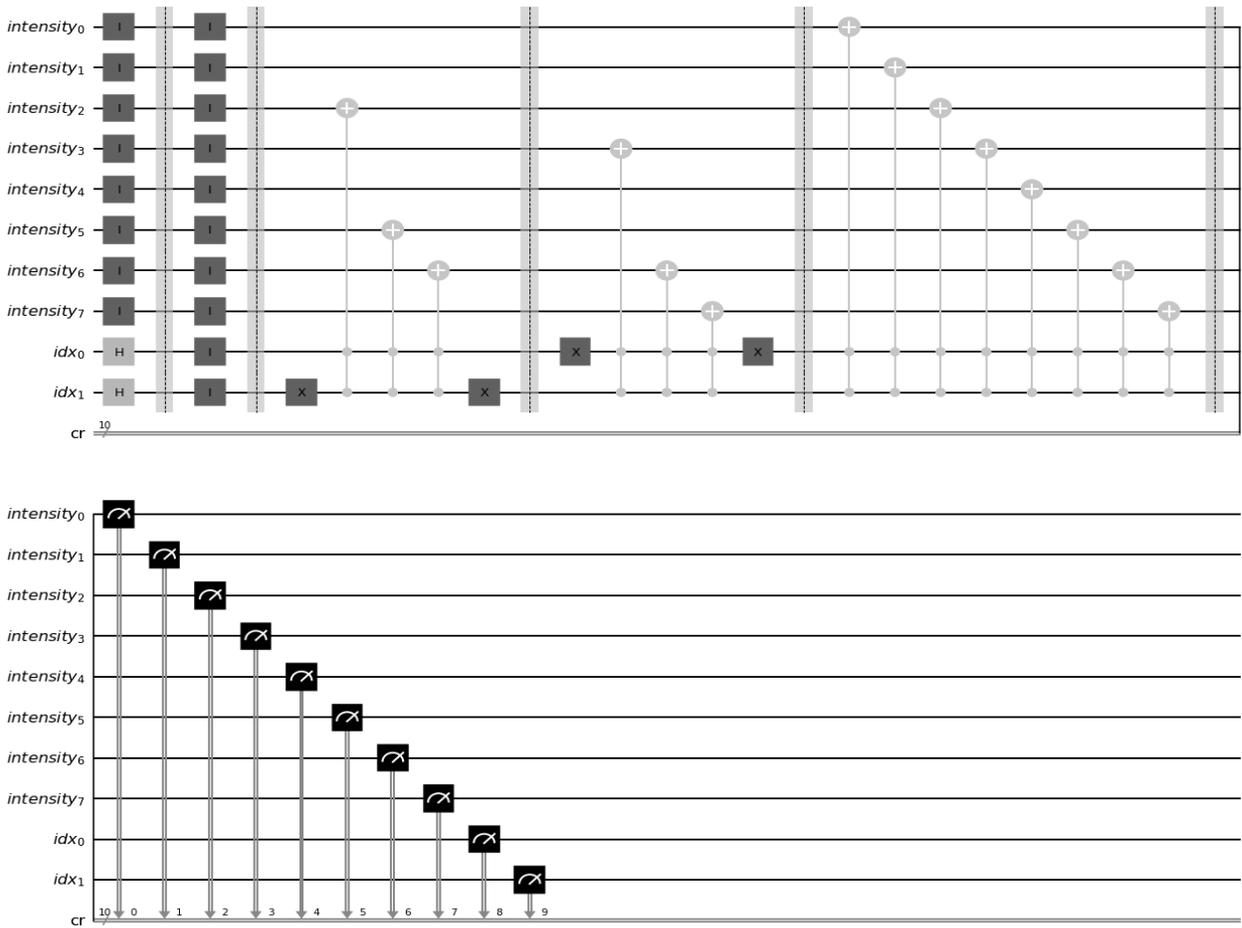

**Fig.1(b)** Quantum Circuit of the above image shown in fig1(a).

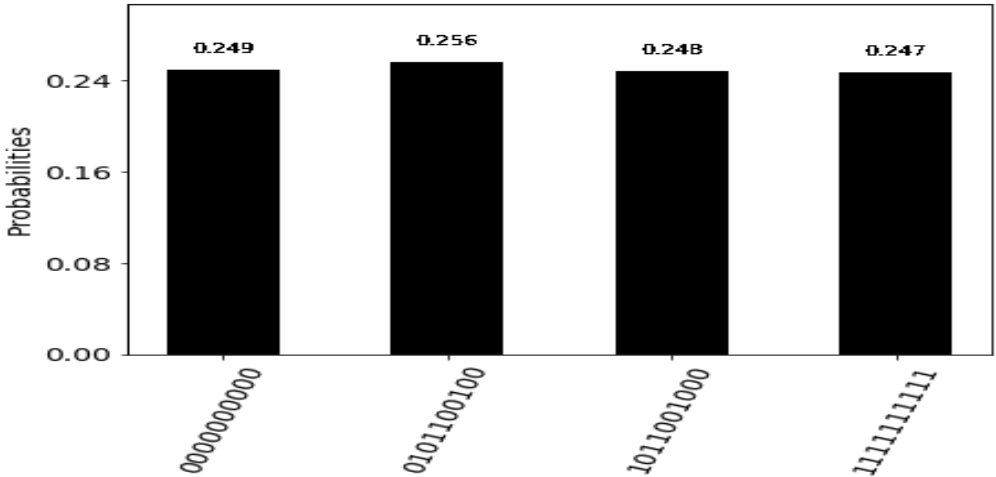

**Fig.1(c)** Histogram showing results of the image after encoding.

In fig.1, (b) first eight qubits (0-7) are used to encode pixel intensities and the other two qubits (8 and 9) is used to encode position information of the pixels. The possible pixel positions are obtained by taking the Hadamard transform of two-position qubits which gives us all the four possible position states. The first pixel in fig.1(a) has intensity 0 which in NEQR is represented by bitstring |00000000⟩. This is encoded by taking eight Identity gates overall intensity Qubits. Similarly, the value setting of other pixels is done by using CC-NOT gates at different positions in the circuit depending on the intensity value of the pixels. At the end of the circuit ten Classical registers are used to measure the output which is shown in fig.1(c). All the pixels are encoded with their respective intensity values. The output of the circuit is a ten-qubit bitstring in which the first two bits represent the position information and the remaining eight bits represent the color information of the pixels. Now our final task is to use Grover algorithm to identify the state with minimum intensity. After running the Grover algorithm for the image, the results are plotted on a histogram as shown in fig.1(d)

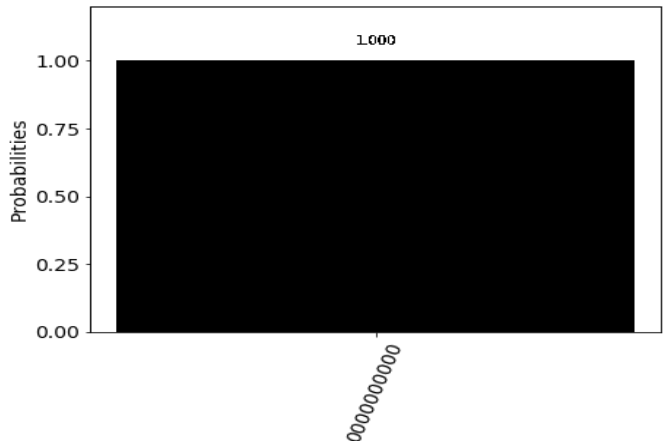

**Fig.1(d)** showing the Histogram results of the Grover algorithm.

**(b) Results based on Google Cirq**

| Sno | Image with intensiy values | Index of black pixel | 3D Histogram |
|---|---|---|---|
| 1 | 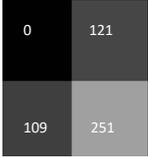 | (0,0) | 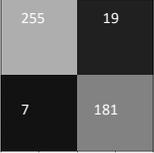 |
| 2 | 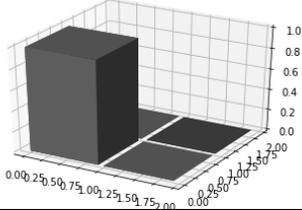 | (0 ,1) | 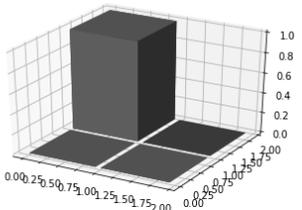 |
| 3 | 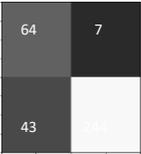 | (1,0) | 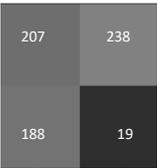 |
| 4 | 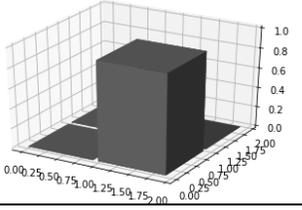 | (1,1) | 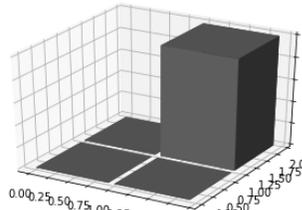 |

Table(2) Different variations of image and their respective plots

**Explanation of the results**

In the above-mentioned table(2) many different variations of intensities for the same image are shown. In first case the intensities of four pixels in a linear array is as [0,121,43,244]. The pixel with the minimum intensity is at location (0,0) i.e it is the first pixel. Similarly, the intensities of all other variations are shown. The results are plotted on a 3d histogram which gives a solid bar of height 1 for the position of minimum intensity pixel and 0 for all other positions.

**Conclusion and future work**

Using quantum mechanics and merging it with image processing is an effective way to address the high computational requirements of classical image processing. In this paper, we have shown two different ways of image processing one based on qiskit and the other on Google cirq. After encoding the image data we aimed to locate the darkest (i.e minimum intensity) pixel among the four pixels using the Grover algorithm. In steganography, we prefer to hide our secret data in the darkest pixel of the image so, that it is not detectable for the intruder. Thus, the location of the black pixel plays an important role and to execute steganography for effective data hiding. It is very much required to know the location of this pixel which can be found with the help of the Grover search algorithm using the exponential speedup processing.

In the future we plan to use this algorithm for image feature detection for higher resolution images by utilizing different image compression schemes to reduce the number of gates being used in the process.